Original Paper

# Improving the Robustness and Clinical Applicability of Automatic Respiratory Sound Classification Using Deep Learning–Based Audio Enhancement: Algorithm Development and Validation

Jing-Tong Tzeng[1], BSc; Jeng-Lin Li[2], PhD; Huan-Yu Chen[2], PhD; Chun-Hsiang Huang[3], MD; Chi-Hsin Chen[3], MD; Cheng-Yi Fan[3], MD; Edward Pei-Chuan Huang[3,4*], MD; Chi-Chun Lee[1,2*], PhD

[1]College of Semiconductor Research, National Tsing Hua University, Hsinchu, Taiwan
[2]Department of Electrical Engineering, National Tsing Hua University, Hsinchu, Taiwan
[3]Department of Emergency Medicine, National Taiwan University Hospital Hsin-Chu Branch, Hsinchu, Taiwan
[4]Department of Emergency Medicine, National Taiwan University Hospital, Taipei, Taiwan
[*]these authors contributed equally

**Corresponding Author:**
Chi-Chun Lee, PhD
Department of Electrical Engineering
National Tsing Hua University
101, Section 2, Kuang-Fu Road
Hsinchu, 300
Taiwan
Phone: 886 35162439
Email: cclee@ee.nthu.edu.tw

**Related Article:**
This is a corrected version. See correction statement in: https://ai.jmir.org/2025/1/e76150

## Abstract

**Background:** Deep learning techniques have shown promising results in the automatic classification of respiratory sounds. However, accurately distinguishing these sounds in real-world noisy conditions poses challenges for clinical deployment. In addition, predicting signals with only background noise could undermine user trust in the system.

**Objective:** This study aimed to investigate the feasibility and effectiveness of incorporating a deep learning–based audio enhancement preprocessing step into automatic respiratory sound classification systems to improve robustness and clinical applicability.

**Methods:** We conducted extensive experiments using various audio enhancement model architectures, including time-domain and time-frequency–domain approaches, in combination with multiple classification models to evaluate the effectiveness of the audio enhancement module in an automatic respiratory sound classification system. The classification performance was compared against the baseline noise injection data augmentation method. These experiments were carried out on 2 datasets: the International Conference in Biomedical and Health Informatics (ICBHI) respiratory sound dataset, which contains 5.5 hours of recordings, and the Formosa Archive of Breath Sound dataset, which comprises 14.6 hours of recordings. Furthermore, a physician validation study involving 7 senior physicians was conducted to assess the clinical utility of the system.

**Results:** The integration of the audio enhancement module resulted in a 21.88% increase with $P<.001$ in the ICBHI classification score on the ICBHI dataset and a 4.1% improvement with $P<.001$ on the Formosa Archive of Breath Sound dataset in multi-class noisy scenarios. Quantitative analysis from the physician validation study revealed improvements in efficiency, diagnostic confidence, and trust during model-assisted diagnosis, with workflows that integrated enhanced audio leading to an 11.61% increase in diagnostic sensitivity and facilitating high-confidence diagnoses.

**Conclusions:** Incorporating an audio enhancement algorithm significantly enhances the robustness and clinical utility of automatic respiratory sound classification systems, improving performance in noisy environments and fostering greater trust among medical professionals.

XSL•FO
RenderX

## Introduction

### Background

Respiratory sounds play a crucial role in pulmonary pathology. They provide insights into the condition of the lungs noninvasively and assist disease diagnosis through specific sound patterns and characteristics [1,2]. For instance, wheezing is a continuous high-frequency sound that often indicates typical symptoms of chronic obstructive pulmonary disease and asthma [3]; crackling, on the other hand, is an intermittent low-frequency sound with a shorter duration that is a common respiratory sound feature among patients with lung infections [4]. The advancement of machine learning algorithms and medical devices enables researchers to investigate approaches for developing automated respiratory sound classification systems, reducing the reliance on manual inputs from physicians and medical professionals.

In earlier studies, researchers have engineered handcrafted audio features for respiratory sound classification [5]. Recently, neural network–based methods have become the de facto methods for lung sound classification. For example, Kim et al [6] fine-tuned the pretrained VGG16 algorithm, outperforming the conventional support vector machine (SVM) classifier. Wanasinghe et al [7] incorporated mel spectrograms, mel-frequency cepstral coefficients, and chroma features to expand the feature set input to a convolutional neural network (CNN), demonstrating promising results in the identification of pulmonary diseases. Pessoa et al [8] proposed a hybrid CNN model architecture that integrates time-domain information with spectrogram-based features, delivering a satisfactory performance. Moreover, various advanced architectures have been proposed to extract both long-term and short-term information from respiratory sounds based on the characteristics of crackle and wheeze sounds and have shown enhanced performance [9-13]. Recent works have used advanced contrastive learning strategies to enhance intraclass compactness and interclass separability for further improvements [14-17]. These advancements in neural network structures have shown increasing promise in achieving reliable respiratory sound classification.

Despite these advancements, significant challenges remain for the clinical deployment of automatic respiratory sound classification systems due to complex real-world noisy conditions [6,18]. Augmentation techniques, such as time shifting, speed tuning, and noise injection, have been key strategies to effectively improve the noise robustness and generalizability of a machine learning model [9,14,16,19-23]. While these approaches have shown promising results in respiratory sound classification tasks, their practical utility as modules for building clinical decision support systems remains in doubt. This is primarily attributed to their inability to provide clinicians with intelligible raw audio to listen to facilitate decision-making, thus making the current augmentation-based approach seem black box and hindering acceptance and adoption by medical professionals.

In fact, given the blooming use of artificial intelligence (AI) in health care, the issue of liability has been the focus. The prevailing public opinion suggests that physicians are the ones to bear responsibility for errors attributed to AI [24]. Hence, when these systems are opaque and inaccessible to physicians, it becomes challenging to have them assume responsibility without a clear understanding of the decision-making process. This difficulty is particularly pronounced for seasoned and senior physicians, who hesitate to endorse AI recommendations without transparent rationale. The resulting lack of trust contributes to conflicts in clinical applications. Therefore, elucidating the decision-making process is crucial to establishing the trust of physicians [25]. Moreover, exceptions are frequent in the field of medicine. For instance, in cases in which bronchioles undergo significant constriction, the wheezing sound may diminish to near silence, a phenomenon referred to as silent wheezing. This intricacy could confound AI systems, necessitating human intervention (ie, listening directly to the recorded audio) [26].

To address these challenges, we propose an approach that involves integrating an audio enhancement module into the respiratory sound classification system, as shown in Figure 1. This module aims to achieve noise-robust respiratory sound classification performance while providing clean audio recordings on file to support physicians' decision-making. By enhancing the audio quality and preserving critical information, our system aimed to facilitate more accurate assessments and foster trust among medical professionals. Specifically, we devised 2 major experiments to evaluate this approach in this study. First, we compared the performance of our noise-robust system through audio enhancement to the conventional method of noise augmentation (noise injection) under various clinical noise conditions and signal-to-noise ratios (SNRs). Second, we conducted a physician validation study to assess confidence and reliability when listening to our cleaned audio for respiratory sound class identification. To the best of our knowledge, this is the first study showing that deep learning enhancement architecture can effectively remove noise while preserving discriminative information for respiratory sound classification algorithms and physicians. Importantly, this study validates the clinical potential and practicality of our proposed audio enhancement front-end module, contributing to more robust respiratory sound classification systems and aiding physicians in making accurate and reliable assessments.

**Figure 1.** An overview of our proposed noise-robust respiratory sound classification system with audio enhancement. CNN: convolutional neural network; CNN14: 14-layer CNN; conformer: convolution-augmented transformer; ISTFT: inverse short-time Fourier transform; STFT: short-time Fourier transform; TS: 2 stage.

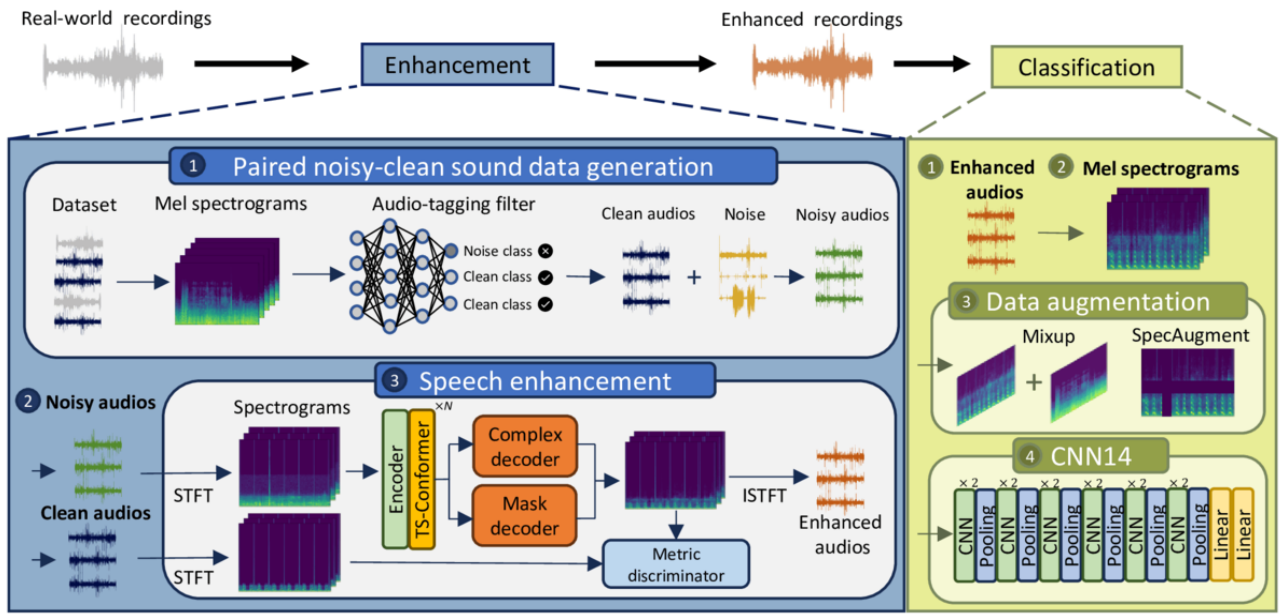

## Related Work

### Audio Enhancement

Audio enhancement is a technique that has been widely used in the speech domain, where it is referred to as speech enhancement. These techniques are primarily used in the front-end stage of automatic speech recognition systems to improve intelligibility [27-29]. Within speech enhancement, deep neural network approaches can be categorized into 2 main domains: time-frequency–domain approaches and time-domain approaches.

Time-frequency–domain approaches are used to estimate clean audio from the short-time Fourier transform (STFT) spectrogram, which provides both time and frequency information. Kumar and Florencio [30] leveraged noise-aware training [31] with psychoacoustic models, which decided the importance of frequency for speech enhancement. The result demonstrated the potential of deep neural network–based speech enhancement in complex multiple-noise conditions, such as real-world environments. In the research by Yin et al [32], they designed a 2-stream architecture that predicts amplitude and phase separately and further improves the performance. However, various research studies [33-35] have indicated that the conventional loss functions used in regression models (eg, $L_1$ and $L_2$) do not strongly correlate with speech quality, intelligibility, and word error rate. To address the issue of discriminator evaluation mismatch, Fu et al [36] introduced MetricGAN. This approach tackles the problem of metrics that are not entirely aligned with the discriminator's way of distinguishing between real and fake samples. They used perceptual evaluation of speech quality (PESQ) [37] and short-time objective intelligibility (STOI) [38] as evaluation functions, which are commonly used for assessing speech quality and intelligibility, as labels for the discriminator. Furthermore, the performance of MetricGAN can be enhanced by adding a learnable sigmoid function for mask estimation, including noisy recording for discriminator training, and using a replay buffer to increase sample size [39]. Recently, convolution-augmented transformers (conformers) have been widely used in automatic speech recognition and speech separation tasks due to their capacity in long-range and local contexts [40-42]. Cao et al [43] introduced a conformer-based metric generative adversarial network (CMGAN), which leverages the conformer structure along with MetricGAN for speech enhancement. In the CMGAN model, multiple 2-stage conformers are used to aggregate magnitude and complex spectrogram information in the encoder. In the decoder, the prediction of the magnitude and complex spectrogram are decoupled and then jointly incorporated to reconstruct the enhanced recordings. Furthermore, CMGAN achieved state-of-the-art results on the VoiceBank+DEMAND dataset [44,45].

On the other hand, time-domain approaches directly estimate the clean audio from the raw signal, encompassing both the magnitude and phase information, enabling them to enhance noisy speech in both domains jointly. Macartney and Weyde [46] leveraged Wave-U-Net, proposed in the study by Thiemann et al [44], to use the U-Net structure in a 1D time domain and demonstrated promising results in audio source separation for speech enhancement. Wave-U-Net uses a series of downsampling and upsampling blocks with skip connections to make predictions. However, its effectiveness in representing long signal sequences is limited due to its restricted receptive field. To overcome this limitation, the approaches presented in the studies by Pandey and Wang [47] and Wang et al [48] divided the signals into small chunks and repeatedly processed local and global information to expand the receptive field. This dual-path structure successfully improved the efficiency in capturing long sequential features. However, dual-path structures are not memory efficient as they require retaining the entire long signal during training. To address the memory efficiency issue, Park et al [49] proposed a multi-view attention network.

They used residual conformer blocks to enrich channel representation and introduced multi-view attention blocks consisting of channel, global, and local attention mechanisms, enabling the extraction of features that reflect both local and global information. This approach also demonstrated state-of-the-art performance on the VoiceBank+DEMAND dataset [44,45].

Both approaches have made significant progress in performance improvements in recent years. However, their suitability for enhancing respiratory sounds collected through stethoscopes remains unclear. Therefore, for this study, we applied these 2 branches of enhancement models and compared their effectiveness in enhancing respiratory sounds in real-world noisy hospital settings [32,43,46,49].

*Respiratory Sound Classification*

In recent years, automatic respiratory sound classification systems have become an active research area. Several studies have explored the use of pretrained weights from deep learning models, showing promising results. Kim et al [6] demonstrated improved performance over SVMs by fine-tuning the pretrained VGG16 algorithm. Gairola et al [22] used effective preprocessing methods, data augmentation techniques, and transfer learning from ImageNet [50] pretrained weights to address data scarcity and further enhance performance.

As large-scale audio datasets [51,52] become more accessible, pretrained audio models are gaining traction, exhibiting promising performance in various audio tasks [53-55]. Studies have explored leveraging these pretrained audio models for respiratory sound classification. Moummad and Farrugia [17] incorporated supervised contrastive loss on metadata with the pretrained 6-layer CNN architecture [53] to improve the quality of learned features from the encoder. Chang et al [56] introduced a novel gamma patch-wise correction augmentation technique, which they applied to the fine-tuned 14-layer CNN (CNN14) architecture [53], achieving state-of-the-art performance. Bae et al [16] used the pretrained Audio Spectrogram Transformer (AST) [54] with a Patch-Mix strategy to prevent overfitting and improve performance. Kim et al [57] proposed a representation-level augmentation technique to effectively leverage different pretrained models with various input types, demonstrating promising results on the pretrained ResNet, EfficientNet, 6-layer CNN, and AST.

However, few of these studies have explicitly addressed the challenge of noise robustness in clinical settings. To improve noise robustness, data augmentation techniques such as adding white noise, time shifting, stretching, and pitch shifting have been commonly used [9,14]. These augmentations enable networks to learn efficient features under diverse recording conditions. Nonetheless, the augmented recordings may not accurately represent the conditions in clinical settings, potentially introducing artifacts and limiting performance improvement. In contrast to the aforementioned works, Kochetov et al [18] proposed a noise-masking recurrent neural network to filter out noisy frames during classification. They concatenated a binary noise classifier and an anomaly classifier with a mask layer to suppress the noisy parts, allowing only the filtered frames to pass through, thereby preventing noises from affecting the classification. However, the International Conference in Biomedical and Health Informatics (ICBHI) database lacks noise labels in the metadata, and the paper did not specify how these labels were obtained, rendering the results nonreproducible. Emmanouilidou et al [58] used multiple noise suppression techniques to address various noise sources, including ambient noise, signal artifacts, heart sounds, and crying, using a soft-margin nonlinear SVM classifier with handcrafted features. Similarly, our work uses a pipeline for noise enhancement and respiratory sound classification. However, we advanced this approach by using deep learning models for both tasks, enabling our system to handle diverse noise types and levels without the need for bespoke strategies for each noise source. Furthermore, we validated our system's practical utility through experiments across 2 respiratory sound databases and a physician validation study, demonstrating its improved performance and clinical relevance.

## Methods

### Datasets

This section presents 2 respiratory sound datasets and 1 clinical noise dataset used in this study.

*ICBHI 2017 Dataset*

The ICBHI 2017 database is one of the largest publicly accessible datasets for respiratory sounds, comprising a total of 5.5 hours of recorded audio [59]. These recordings were independently collected by 2 research teams in Portugal and Greece from 126 participants of all ages (79 adults, 46 children, and 1 unknown). The data acquisition process involved heterogeneous equipment and included recordings from both clinical and nonclinical environments. The duration of the recorded audio varies from 10 to 90 seconds. Within this database, 6898 respiratory cycles result in 920 annotated audio samples. Among these samples, 1864 contain crackles, 886 contain wheezes, and 506 include both crackles and wheezes, whereas the remaining cycles are categorized as normal.

*Formosa Archive of Breath Sound*

The Formosa Archive of Breath Sound (FABS) database comprises 14.6 hours of respiratory sound recordings collected from 1985 participants. Our team collected these recordings at the emergency department of the Hsin-Chu Branch at the National Taiwan University Hospital (NTUH). We used the CaRDIaRT DS101 electronic stethoscope, where each recording is 10 seconds long.

To ensure the accuracy of the annotations, a team of 7 senior physicians meticulously annotated the audio samples. The annotations focused on identifying coarse crackles, wheezes, or normal respiratory sounds. Unlike the ICBHI 2017 database, our annotation process treated each audio sample in its entirety rather than splitting it into respiratory cycles. This approach reduces the need for extensive segmentation procedures and aligns with regular clinical practice. To enhance the quality of the annotations, we implemented an annotation validation flow called "cross-annotator model validation." This involved training multiple models based on each annotator's data and validating the models on data from other annotators. Any data with incongruent predictions were initially identified. These data then underwent additional annotation by 3 senior physicians randomly selected from the original annotation team for each sample to achieve the final consensus label. The FABS database encompasses 5238 annotated recordings, with 715 containing coarse crackles, 234 containing wheezes, and 4289 labeled as normal respiratory sound recordings. The detailed comparison between the ICBHI 2017 dataset and the FABS database is shown in Table 1.

Table 1. Comparison between the International Conference in Biomedical and Health Informatics (ICBHI) and Formosa Archive of Breath Sound (FABS) datasets.

|  | ICBHI (n=126 patients) | FABS (n=1985 patients) |
| --- | --- | --- |
| Age (y), mean (SD) | 42.99 (32.08) | 66.04 (17.64) |
| BMI (kg/m$^2$), mean (SD) | 27.19 (5.34) | 23.95 (4.72) |
| **Sex, n (%)** |  |  |
| Male | 79 (62.7) | 974 (49.1) |
| Female | 46 (36.5) | 841 (42.4) |
| Unknown | 1 (0.8) | 170 (8.6) |
| Sampling rate (kHz) | 4-44.1 | 16 |
| Duration (hours) | 5.5 | 14.6 |
| Label | Crackle and wheeze, crackle, wheeze, and normal | Coarse crackle, wheeze, and normal |
| Equipment | AKG C417L microphone, Littmann Classic II SE stethoscope, Littmann 3200 electronic stethoscope, and Welch Allyn Meditron electronic stethoscope | CaRDIaRT DS101 electronic stethoscope |

### *NTUH Clinical Noise Dataset*

The noise dataset used in this study was sourced from the NTUH Hsin-Chu Branch. To replicate the noise sounds that physicians typically encounter in real-world clinical settings, we used the CaRDIaRT DS101 electronic stethoscope for collecting the noise samples. The NTUH clinical noise dataset consists of 3 different types of clinical noises: 8 friction noises produced by the stethoscope moving on different fabric materials; 18 environment noises recorded at various locations within the hospital; and 12 patient noises generated by patients during auscultation through conversations, coughing, and snoring.

### **Proposed Methods**

As shown in Figure 1, our proposed noise-robust respiratory sound classification system includes two main components: (1) audio enhancement and (2) respiratory sound classifier.

### *Audio Enhancement Module*

Audio enhancement is usually approached as a supervised learning problem [30,31,33-36,39,43], where the goal is to map noisy respiratory sound inputs to their clean counterparts. Mathematically, this task can be represented as learning a function $f$, mapping $X_{noisy}$ to $X_{clean}$, where $X_{noisy}$ represents the input noisy sound and $X_{clean}$ denotes the corresponding clean sound. The enhanced output, $X'_{clean}$, is obtained as $X'_{clean}=f(X_{noisy})$ (1), where $f$ is the audio enhancement model optimized during training.

To achieve high-quality enhancement, it is crucial to carefully select reference clean recordings from the respiratory sound database to generate high-quality paired noisy-clean sound data. To address this, we used an "audio-tagging filter" approach. This approach leverages a large pretrained audio-tagging model to identify clean samples and exclude recordings with irrelevant tags from the database. Specifically, we used the CNN14 pretrained audio neural network [53] that was trained on AudioSet [51], a comprehensive audio dataset containing 2,063,839 training audio clips sourced from YouTube covering 527 sound classes. Audio samples with the following audio event labels were filtered out: "music," "speech," "fire," "animal," "cat," and "domestic animals, pets." These labels were chosen as they were among the top commonest predictions of the audio-tagging model, indicating a higher likelihood of significant irrelevant noise in the recordings. By excluding these labels, we could ensure that the selected recordings could be used as reference clean audio. To validate the effectiveness of the filtering process, we manually checked the filtered recordings. The results showed that the tagging precision was 92.5%, indicating that this method is efficient and trustworthy. Moreover, as it is fully automatic, it is easy to reproduce the results.

In the ICBHI 2017 database, 889 clean audio samples were retained after filtering, consisting of 1812 cycles with crackling sounds, 822 cycles with wheezing sounds, 447 cycles with both crackling and wheezing sounds, and 3538 cycles with normal respiratory sounds. Alternatively, the filtered FABS clean samples comprised 699 recordings of coarse crackle respiratory sounds, 225 recordings of wheeze respiratory sounds, and 4238 recordings of normal respiratory sounds.

In this study, we used Wave-U-Net [46], Phase-and-Harmonics–Aware Speech Enhancement Network (PHASEN) [32], Multi-View Attention Network for Noise Erasure [49], and CMGAN [43] to compare the effectiveness of different model structures in enhancing respiratory sounds.

## Respiratory Sound Classification

Training a classification model from scratch using a limited dataset may lead to suboptimal performance or overfitting. Therefore, we selected the CNN14 model proposed in the study by Kong et al [53], which had been pretrained on AudioSet [51], as our main classification backbone, and we further fine-tuned it on our respiratory datasets. We used log-mel spectrograms as the input feature, similar to previous works in respiratory sound classifications [6,9-11,14]. As the dataset is highly imbalanced, we used the balanced batch-learning strategy. To further improve model generalizability and performance, we incorporated data augmentation techniques, including Mixup [60] and SpecAugment [61], along with triplet loss [15,62] to enhance feature separability.

Mathematically, the classification task is formulated as a multi-class classification problem. The goal is to learn a mapping function, $g: Z \rightarrow Y$ (2), where $Z$ represents the extracted features and $Y$ denotes the target class labels. To obtain $Z$, input-enhanced audio signals $X'_{clean}$ are transformed using the STFT to generate a spectrogram, followed by mel-filter banks to convert the frequency scale to the mel scale: $Z=\log\text{-mel}(\text{STFT}[X'_{clean}])$ (3).

During training, the total loss function $L_c$ combines cross-entropy loss and triplet loss: $L_c = L_{CE} + \lambda L_{triplet}$ (4).

Through grid search, $\lambda=0.01$ leads to the best performance.

## Physician Validation Study

To further evaluate the effectiveness of audio enhancement for respiratory sound, we conducted a physician validation study using the clean, noisy, and enhanced recordings from a randomly selected 25% of the testing set on the ICBHI 2017 database. In this study, we invited 7 senior physicians to independently annotate these recordings without access to any noise level or respiratory sound class label. We instructed the physicians to label the respiratory class with a confidence score ranging from 1 to 5. The objective was to demonstrate that our proposed method not only enhances the performance of the classification model but also improves the accuracy of the respiratory sound classification and increases the confidence in manual judgment done by physicians. The physician validation study was a critical step in validating the clinical practicality and effectiveness of our proposed audio enhancement preprocessing technique in clinical settings.

## Ethical Considerations

This study was approved by the institutional review board of the NTUH Hsin-Chu Branch (109-129-E) and complies with ethical guidelines for human research. It involved both prospective and retrospective data collection, with retrospective data fully deidentified to protect participant privacy. All prospective participants provided informed consent before data collection. No financial compensation was provided to participants, ensuring voluntary and unbiased participation.

# Results

## Overview

To assess the noise robustness of our proposed method, we conducted a comparative analysis using methods across various levels of noise intensity, as outlined in Textbox 1.

**Textbox 1.** Methods for various levels of noise intensity.

| |
|---|
| **Clean** |
| The respiratory sound classification models were only trained on clean data and tested on clean data. This approach served to establish the upper-bound performance for the overall comparison. |
| **Noisy** |
| The respiratory sound classification models were trained on clean data but tested on noisy data. As the models were not optimized for noise robustness, a significant drop in performance was expected. |
| **Noise injection** |
| The respiratory sound classification models were trained on synthesized noisy data and tested on noisy data. This approach represents the conventional method to enhance the noise robustness of the model. |
| **Audio enhancement** |
| The audio enhancement model functions as a front-end preprocessing step for the classification model. To achieve this, we first optimized the audio enhancement model to achieve a satisfactory enhancement performance. Subsequently, the respiratory sound classification model was trained on the enhanced data and tested on the enhanced data. |

## Experiment Setup

To evaluate the efficiency of our proposed method, we followed a similar setup as that in prior work [6,11,14] to have an 80%-20% train-test split on the database. Furthermore, the training set was mixed with the noise recordings from the NTUH clinical noise dataset with 4 SNRs (15, 10, 5, and 0 dB) with random time shifting. The test set was mixed with unseen noise data with 4 SNRs (17.5, 12.5, 7.5, and 2.5 dB), also subjected to random time shifting. For evaluation, we used the metrics of accuracy, sensitivity, specificity, and ICBHI score. Sensitivity is defined as the recall of abnormal respiratory sounds. Specificity refers to the recall of normal respiratory sounds. The ICBHI score, calculated as the average of sensitivity and specificity, provides a balanced measure of the model's classification performance.

## Implementation Details

### Technical Setup

The models were implemented using PyTorch (version 1.12; Meta AI) with the CUDA Toolkit (version 11.3; NVIDIA Corporation) for graphics processing unit acceleration. Training was conducted on an NVIDIA A100 graphics processing unit with 80 GB of memory. For clarity and reproducibility, the detailed implementation and computational setup is provided in Multimedia Appendix 1.

### Preprocessing

We first resampled all recordings to 16 kHz. Next, each respiratory cycle was partitioned into 10-second audio segments before proceeding with feature extraction. In cases in which cycles were shorter in duration, we replicated and concatenated them to form 10-second clips in the ICBHI dataset. As the recordings in the FABS dataset are initially labeled per recording, there was no requirement for a segmentation process. Subsequently, these audio clips were mixed with the NTUH clinical noise dataset, generating pairs of noisy and clean data for further processing.

### Enhancement Model Training

For enhancement model training, the 10-second audio clips were divided into 4-second segments. When implementing Wave-U-Net [43], the channel size was set to 24, the batch size was set to 4, and the number of layers of convolution upsampling and downsampling was set to 8. The model was trained using the Adam optimizer with a learning rate of $10^{-5}$ for 40 epochs when training using pretrained weights and $10^{-4}$ for 30 epochs when training from scratch. For the Multi-View Attention Network for Noise Erasure model [49], the channel size was set to 60, the batch size was set to 4, and the number of layers of up and down convolution was set to 4. The model was trained using the Adam optimizer with a learning rate of $10^{-6}$ for 10 epochs when training using pretrained weights and a learning rate of $10^{-5}$ for 10 epochs when training from scratch. When implementing PHASEN [32], which is trained in the time-frequency domain, we followed the original setup using a Hamming window of 25 ms in length and a hop size of 10 ms to generate STFT spectrograms. The number of 2-stream blocks was set to 3, the batch size was set to 4, the channel number for the amplitude stream was set to 24, and the channel number for the phase stream was set to 12. The model was trained using the Adam optimizer with a learning rate of $5 \times 10^{-5}$ for 20 epochs when training using pretrained weights and a learning rate of $5 \times 10^{-4}$ for 30 epochs when training from scratch. For CMGAN [43], we followed the original setting using a Hamming window of 25 ms in length and a hop size of 6.25 ms to generate STFT spectrograms. The number of 2-stage conformer blocks was set to 4, the batch size was set to 4, and the channel number in the generator was set to 64. The channel numbers in the discriminator were set to 16, 32, 64, and 128. The model was trained using the Adam optimizer with a learning rate of $5 \times 10^{-5}$ for 20 epochs when training using pretrained weights and a learning rate of $5 \times 10^{-4}$ for 30 epochs when training from scratch. These hyperparameters are also listed in Multimedia Appendix 2.

The pretrained weights for these models were trained on the VoiceBank+DEMAND dataset [44,45], which is commonly used in speech enhancement research.

### Classification Model Training

For the classification model, the 4-second enhanced segments were concatenated back into 10-second audio clips. To generate the log-mel spectrogram, the waveform was transformed using STFT with a Hamming window size of 512 and a hop size of 160 samples. The STFT spectrogram was then processed through 64 mel filter banks to generate the log-mel spectrogram. In the training stage, we set the batch size to 32 and used the Adam optimizer with a learning rate of $10^{-4}$ for 14,000 iterations using pretrained weights from the model trained on the 16-kHz AudioSet dataset [51]. These hyperparameters are also listed in Multimedia Appendix 2.

## Evaluation Outcomes

In this study, we compared the classification performance of conventional noisy data augmentation with our proposed audio-enhanced preprocessing. The test set was split into 2 groups, and each classification model was trained 10 times, yielding 20 values for statistical analysis. We conducted a 1-tailed *t* test to assess whether models trained on CMGAN-enhanced audio using pretrained weights showed significant improvements over other models. In addition, we reported speech quality metrics for various audio enhancement models and analyzed their correlation with classification performance.

The experiment results, as shown in Table 2, highlight the effectiveness of our proposed audio enhancement preprocessing strategy for noise-robust performances. In the case of the ICBHI 2017 database, the model trained solely on clean data experienced a 33.95% drop in the ICBHI score when evaluated on the synthesized noisy dataset. Noise injection improved the score by 19.73%, but fine-tuning PHASEN achieved the highest score, outperforming noise injection by 2.28%. Regarding the FABS database, using the classification model trained on clean recordings on the noisy recordings led to a 12.48% drop in the ICBHI score. Noise injection improved performance by 1.31%, but fine-tuning CMGAN outperformed noise injection by 2.79%. Across both datasets, the audio enhancement preprocessing method consistently improved performance compared to the noise injection augmentation technique. Furthermore, it showed improved sensitivity for all enhancement model structures, with the most significant improvement being 6.31% for the ICBHI database and 13.54% for the FABS database. This indicates that the audio enhancement preprocessing method enhanced the classification model's ability to distinguish abnormal respiratory sounds, which is crucial for the early detection of potential illnesses in clinical use.

**Table 2.** Comparison of classification performance on both the International Conference in Biomedical and Health Informatics (ICBHI) and Formosa Archive of Breath Sound (FABS) datasets.

| Method | Enhancement model | Accuracy, mean (SD) | P value | Sensitivity, mean (SD) | P value | Specificity, mean (SD) | P value | ICBHI score, mean (SD) | P value |
|---|---|---|---|---|---|---|---|---|---|
| **ICBHI** | | | | | | | | | |
| Clean | —[a] | 79.90 (0.01) | >.99 | 71.43 (0.02) | >.99 | 87.27 (0.01) | >.99 | 79.35 (0.01) | >.99 |
| Noisy | — | 45.70 (0.03) | <.001 | 40.99 (0.04) | <.001 | 49.80 (0.08) | <.001 | 45.40 (0.03) | <.001 |
| Noise injection | — | 65.85 (0.01) | <.001 | 54.89 (0.04) | <.001 | 75.37 (0.04) | .98 | 65.13 (0.01) | <.001 |
| AE[b] | Wave-U-Net | 60.86 (0.02) | <.001 | 55.35 (0.04) | <.001 | 65.66 (0.05) | <.001 | 60.50 (0.02) | <.001 |
| AE | Wave-U-Net[c] | 61.29 (0.02) | <.001 | 55.04 (0.02) | <.001 | 66.72 (0.04) | <.001 | 60.88 (0.02) | <.001 |
| AE | PHASEN[d] | 66.81 (0.01) | .02 | 57.61 (0.03) | .001 | 74.81 (0.04) | .91 | 66.21 (0.01) | .005 |
| AE | PHASEN[c] | 68.09[e] (0.01) | .84 | 57.71[f] (0.03) | .004 | 77.12[f] (0.04) | >.99 | 67.41[e] (0.01) | .64 |
| AE | MANNER[g] | 67.62 (0.01) | .39 | 53.09 (0.03) | <.001 | 80.26[e] (0.04) | >.99 | 66.67 (0.01) | .03 |
| AE | MANNER[c] | 60.36 (0.02) | <.001 | 57.67 (0.02) | <.001 | 62.70 (0.04) | <.001 | 60.19 (0.02) | <.001 |
| AE | CMGAN[h] | 64.75 (0.01) | <.001 | 55.84 (0.03) | <.001 | 72.50 (0.02) | .17 | 64.17 (0.01) | <.001 |
| AE | CMGAN[c] | 67.70[f] (0.01) | — | 61.20[e] (0.03) | — | 73.35 (0.02) | — | 67.28[f] (0.01) | — |
| **FABS** | | | | | | | | | |
| Clean | — | 85.02 (0.01) | >.99 | 62.07 (0.04) | >.99 | 90.01 (0.02) | <.001 | 76.04 (0.02) | >.99 |
| Noisy | — | 81.02 (0.02) | <.001 | 36.41 (0.04) | <.001 | 90.71 (0.02) | .004 | 63.56 (0.02) | <.001 |
| Noise injection | — | 84.53 (0.01) | >.99 | 34.29 (0.05) | <.001 | 95.44 (0.01) | >.99 | 64.87 (0.02) | <.001 |
| AE | Wave-U-Net | 85.97[e] (0.01) | >.99 | 36.74 (0.03) | <.001 | 96.66[f] (0.01) | >.99 | 66.70 (0.01) | .04 |
| AE | Wave-U-Net[c] | 85.88[f] (0.01) | >.99 | 29.08 (0.05) | <.001 | 98.22[e] (0.01) | >.99 | 63.65 (0.02) | <.001 |
| AE | PHASEN | 85.29 (0.004) | >.99 | 33.64 (0.02) | <.001 | 96.51 (0.01) | >.99 | 65.07 (0.01) | <.001 |
| AE | PHASEN[c] | 85.33 (0.01) | >.99 | 35.82 (0.03) | <.001 | 96.09 (0.01) | >.99 | 65.95 (0.02) | <.001 |
| AE | MANNER | 83.01 (0.01) | .05 | 37.50 (0.08) | .01 | 92.89 (0.03) | .67 | 65.20 (0.03) | .004 |
| AE | MANNER[c] | 79 (0.03) | <.001 | 47.83[e] (0.06) | >.99 | 85.77 (0.05) | <.001 | 66.80[f] (0.02) | .08 |
| AE | CMGAN | 82.47 (0.01) | <.001 | 37.61 (0.05) | <.001 | 92.22 (0.01) | .19 | 64.91 (0.02) | <.001 |
| AE | CMGAN[c] | 83.67 (0.01) | — | 42.77[f] (0.03) | — | 92.55 (0.01) | — | 67.66[e] (0.01) | — |

[a]Without any audio enhancement module.

[b]AE: audio enhancement.
[c]The model is fine-tuned from the pretrained weight.
[d]PHASEN: Phase-and-Harmonics–Aware Speech Enhancement Network.
[e]Best performance across all methods for this metric.
[f]Second-best performance across all methods for this metric.
[g]MANNER: Multi-View Attention Network for Noise Erasure.
[h]CMGAN: convolution-augmented transformer–based metric generative adversarial network.

Comparing the 2 types of enhancement approaches, the time-frequency domain models (PHASEN and CMGAN) exhibited better performance in terms of ICBHI scores. In addition, CMGAN consistently showed high sensitivity across both datasets, indicating its potential for preserving respiratory sound features during audio enhancement. The spectrogram of the audio enhanced using CMGAN also revealed that it preserves more high-frequency information across all respiratory sound classes, as illustrated in Figure 2. In contrast, audio enhanced using other models either lost high-frequency information or retained too much noise, leading to misclassification as normal, resulting in higher specificity for those models. Moreover, we observed that, while our focus was on training a respiratory sound enhancement model, using pretrained weights from models trained on the VoiceBank+DEMAND dataset, which were originally designed for speech, still significantly improved classification performance in most cases. This highlights the cross-domain effectiveness of pretrained weights from the speech domain in respiratory sound tasks.

**Figure 2.** The log-mel spectrograms of 4 different types of respiratory sounds on the International Conference in Biomedical and Health Informatics 2017 database. Each subfigure contains clean audio, noisy audio, and 4 types of enhanced audio from different audio enhancement approaches. CMGAN: convolution-augmented transformer–based metric generative adversarial network; MANNER: Multi-View Attention Network for Noise Erasure; PHASEN: Phase-and-Harmonics–Aware Speech Enhancement Network.

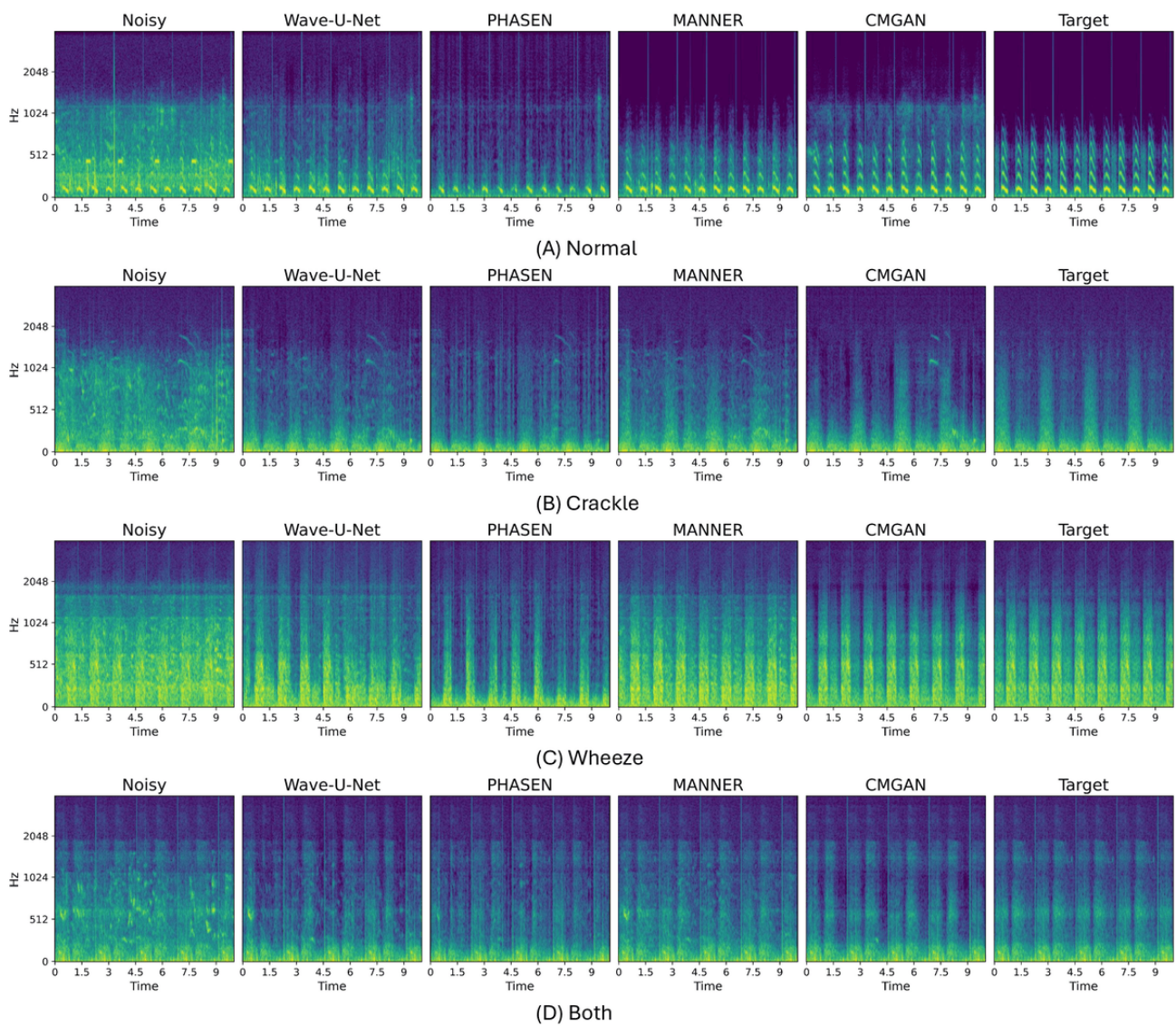

To evaluate whether speech quality metrics, originally designed for speech, are effective for respiratory sounds, we analyzed their correlation with the ICBHI score and sensitivity. As shown in Table 3, the mean opinion score (MOS) of background noise intrusiveness (CBAK) and segmental SNR (SSNR) exhibited relatively higher correlations than other metrics, such as PESQ, STOI, the MOS of signal distortion, and the MOS of overall quality. Unlike these other metrics, which are primarily designed to assess speech intelligibility and quality, CBAK and SSNR focus on background noise intrusiveness and the SNR between recordings. This distinction explains why CBAK and SSNR show stronger correlations with classification performance, highlighting their potential applicability for respiratory sound analysis.

We evaluated the inference times of 4 audio enhancement models. Wave-U-Net generates 1 second of enhanced audio in just 1.5 ms, PHASEN does so in 3.9 ms, and MANNER does so in 11.7 ms. In contrast, CMGAN processes 1 second of audio in 26 ms—a longer time that is offset by its superior classification performance.

To further analyze the effectiveness of our proposed audio enhancement preprocessing method in handling different types of noise, we compared its performance using the noise injection method across various SNR levels. On the basis of the consistently outstanding performance of CMGAN across both datasets, we selected it for further analysis.

On the ICBHI database, as illustrated in Figure 3, the noise injection method performed better with environmental noises at SNR values of 2.5 and 12.5 dB. However, the front-end audio enhancement consistently performed better for patient and friction noises across almost all noise levels.

Regarding the FABS dataset, as shown in Figure 4, the noise injection method performed better with environmental and friction noises at an SNR value of 17.5 dB and patient noises at an SNR value of 2.5 and 7.5 dB. In all other situations, the audio enhancement preprocessing method demonstrated superior ICBHI scores.

These results suggest that our proposed strategy effectively mitigates the effects of various noise types while maintaining strong classification performance. This highlights the robustness and reliability of our approach in handling diverse noise scenarios and intensities, showcasing its potential for practical applications in clinical settings.

**Table 3.** Comparison of audio enhancement (AE) performance on both the International Conference in Biomedical and Health Informatics (ICBHI) and Formosa Archive of Breath Sound (FABS) datasets.

| Method | Enhancement model | Parameters (millions) | PESQ[a,b] | CSIG[c,d] | CBAK[e,f] | COVL[g,h] | SSNR[i,j] | STOI[k,l] |
|---|---|---|---|---|---|---|---|---|
| **ICBHI** | | | | | | | | |
| Noisy | —[m] | — | 0.58 | 2.98 | 2.83 | 2.13 | 14.10 | 0.50 |
| AE | Wave-U-Net | 3.3 | 0.56 | 3.07 | 3.25 | 2.18 | 20.30 | 0.49 |
| AE | Wave-U-Net[n] | 3.3 | 0.57 | 3.10 | 3.25 | 2.20 | 20.20 | 0.50 |
| AE | PHASEN[o] | 7.7 | 0.57 | 3.07 | 3.34 | 2.19 | 21.41 | 0.52 |
| AE | PHASEN[n] | 7.7 | 0.56 | 3.04 | 3.32 | 2.17 | 21.26 | 0.51 |
| AE | MANNER[p] | 24 | 0.59 | 3.23 | 3.24 | 2.27 | 19.85 | 0.55 |
| AE | MANNER[n] | 24 | 0.66 | 3.38[q] | 3.24 | 2.39[r] | 19.17 | 0.60[r] |
| AE | CMGAN[s] | 1.8 | 0.75[q] | 3.31[r] | 3.46[r] | 2.40[q] | 22.06[r] | 0.61[q] |
| AE | CMGAN[n] | 1.8 | 0.74[r] | 3.29 | 3.47[q] | 2.38 | 22.31[q] | 0.61[q] |
| **FABS** | | | | | | | | |
| Noisy | — | — | 2.10 | 3.80[q] | 3.41 | 3.03[q] | 12.99 | 0.62[r] |
| AE | Wave-U-Net | 3.3 | 1.78 | 1.96 | 3.16 | 1.90 | 10.97 | 0.52 |
| AE | Wave-U-Net[n] | 3.3 | 1.75 | 1.89 | 3.13 | 1.86 | 10.74 | 0.50 |
| AE | PHASEN | 7.7 | 1.93 | 2.34 | 3.26 | 2.19 | 11.54 | 0.58 |
| AE | PHASEN[n] | 7.7 | 1.84 | 2.11 | 3.20 | 2.03 | 11.27 | 0.57 |
| AE | MANNER | 24 | 2.14[r] | 3.35 | 3.44[r] | 2.81 | 12.87 | 0.61 |
| AE | MANNER[n] | 24 | 2.18[q] | 3.57[r] | 3.44[r] | 2.95[r] | 12.57 | 0.63[q] |
| AE | CMGAN | 1.8 | 2.01 | 1.79 | 3.42 | 1.96 | 13.59[r] | 0.59 |
| AE | CMGAN[n] | 1.8 | 2.06 | 1.68 | 3.48[q] | 1.91 | 13.98[q] | 0.59 |

[a]PESQ: perceptual evaluation of speech quality.

[b]ICBHI: sensitivity correlation coefficient=0.36 and ICBHI score correlation coefficient=0.23; FABS: sensitivity correlation coefficient=0.72 and ICBHI score correlation coefficient=0.16.

[c]CSIG: mean opinion score (MOS) of signal distortion.

[d]ICBHI: sensitivity correlation coefficient=0.51 and ICBHI score correlation coefficient=0.40; FABS: sensitivity correlation coefficient=0.34 and ICBHI score correlation coefficient=–0.25.

[e]CBAK: MOS of background noise intrusiveness.

[f]ICBHI: sensitivity correlation coefficient=0.92 and ICBHI score correlation coefficient=0.90; FABS: sensitivity correlation coefficient=0.71 and ICBHI score correlation coefficient=0.23.

[g]CVOL: MOS of overall quality.

[h]ICBHI: sensitivity correlation coefficient=0.52 and ICBHI score correlation coefficient=0.39; FABS: sensitivity correlation coefficient=0.42 and ICBHI score correlation coefficient=–0.20.

[i]SSNR: segmental signal-to-noise ratio.

[j]ICBHI: sensitivity correlation coefficient=0.92 and ICBHI score correlation coefficient=0.93; FABS: sensitivity correlation coefficient=0.59 and ICBHI score correlation coefficient=0.22.

[k]STOI: short-time objective intelligibility.

[l]ICBHI: sensitivity correlation coefficient=0.45 and ICBHI score correlation coefficient=0.36; FABS: sensitivity correlation coefficient=0.68 and ICBHI score correlation coefficient=0.13.

[m]Without any audio enhancement module.

[n]The model is fine-tuned from the pretrained weight.

[o]PHASEN: Phase-and-Harmonics–Aware Speech Enhancement Network.

[p]MANNER: Multi-View Attention Network for Noise Erasure.

[q]Best performance across all methods for this metric.

[r]Second-best performance across all methods for this metric.
[s]CMGAN: convolution-augmented transformer–based metric generative adversarial network.

**Figure 3.** Performance comparison of different approaches for each noise type with various signal-to-noise ratio (SNR) values on the International Conference in Biomedical and Health Informatics (ICBHI) 2017 database.

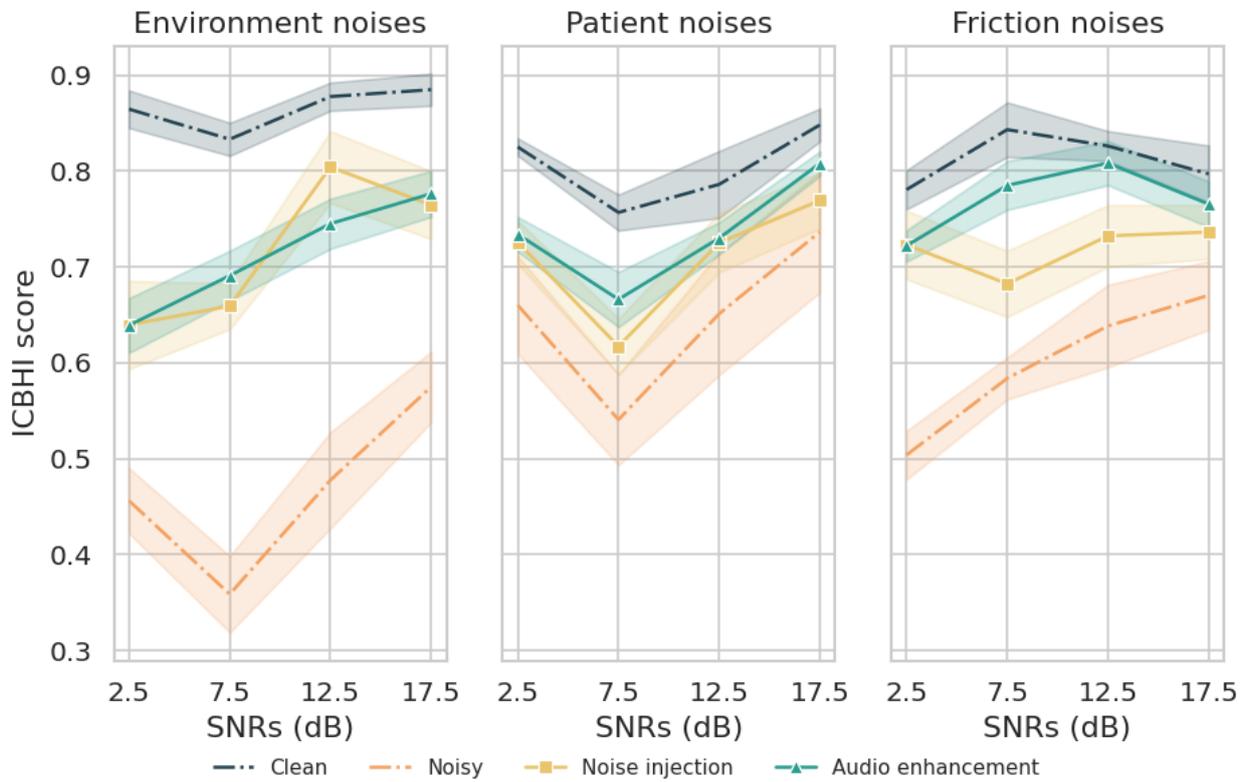

**Figure 4.** Performance comparison of different approaches for each noise type with various signal-to-noise ratio (SNR) values on the Formosa Archive of Breath Sound database. ICBHI: International Conference in Biomedical and Health Informatics.

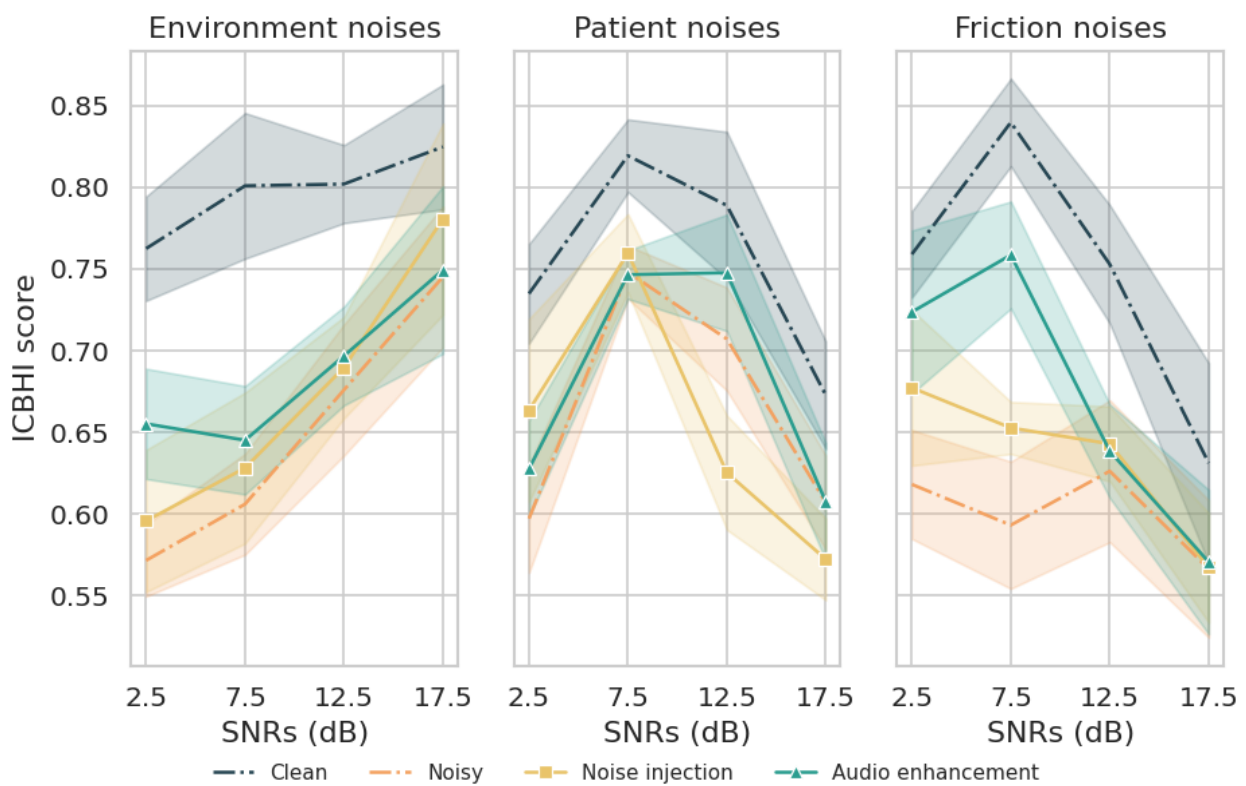

## Physician Validation Study

To assess the practical utility of our proposed approach in clinical settings, we conducted a physician validation study using the ICBHI dataset. This study involved comparing the annotation results provided by 7 senior physicians under 3 different conditions: clean, noisy, and enhanced recordings. By evaluating physician assessments across these conditions, we aimed to determine the effectiveness of our enhancement approach in improving diagnostic accuracy and confidence.

As shown in Table 4, the presence of noise in the recordings had a noticeable impact on the physicians' ability to conduct a reliable judgment, reducing accuracy by 1.81% and sensitivity by 6.46% compared to the clean recordings. However, the recordings with audio enhancement exhibited notable improvement, with a 3.92% increase in accuracy and an 11.61% increase in sensitivity compared to the noisy recordings. The enhanced audio successfully preserved sound characteristics crucial for physicians in classifying respiratory sounds, leading to higher true positive rates in distinguishing adventitious sounds.

The enhanced audio recordings also received higher annotation confidence scores than the noisy recordings, as indicated in Figure 5 and Table 4. Moreover, the speech quality metrics PESQ, MOS of signal distortion, CBAK, MOS of overall quality, SSNR, and STOI positively correlated with the physicians' annotation confidence, as shown in Figure 6. These results underscore the potential of audio enhancement preprocessing techniques for practical application in real-world clinical settings.

**Table 4.** Annotation results from physicians on different types of recordings.

| Type of recording | Accuracy (%) | Sensitivity (%) | Specificity (%) | ICBHI[a] score (%) | Confidence mean (SD) |
| --- | --- | --- | --- | --- | --- |
| Clean | 49.4 | 23.23 | 72.32 | 47.77 | 2.88 (1.50) |
| Noisy | 47.59 | 16.77 | 74.58 | 45.68 | 2.32 (1.29) |
| Enhanced | 51.51 | 28.38 | 71.75 | 50.07 | 2.65 (1.36) |

[a]ICBHI: International Conference in Biomedical and Health Informatics.

**Figure 5.** Physicians' annotation confidence score comparison among clean, noisy, and enhanced recordings.

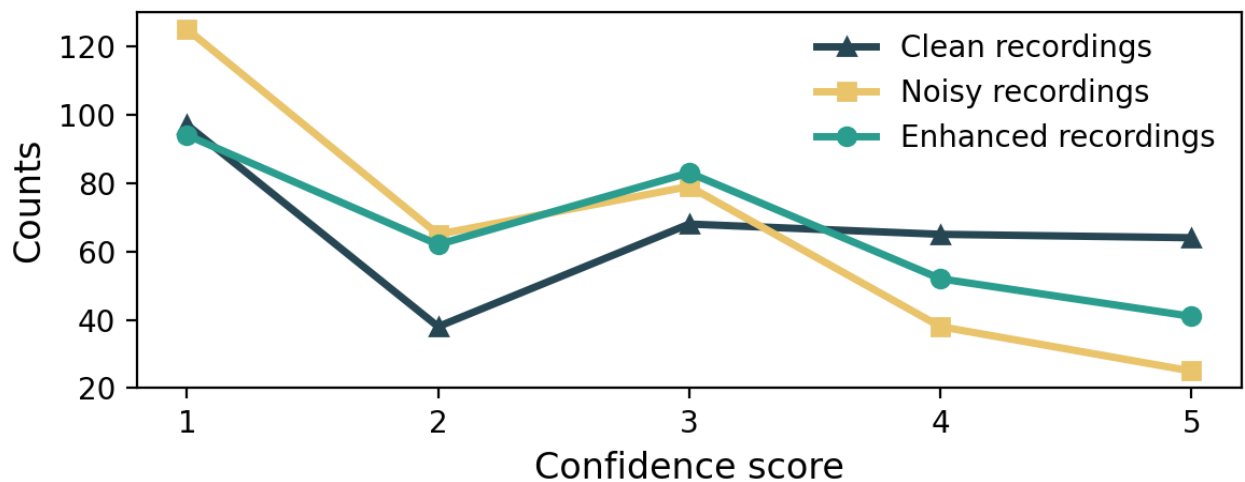

**Figure 6.** Relationship between physicians' annotation confidence score and speech quality metrics. CBAK: mean opinion score (MOS) of background noise intrusiveness; CSIG: MOS of signal distortion; CVOL: MOS of overall quality; PESQ: perceptual evaluation of speech quality; SSNR: segmental signal-to-noise ratio; STOI: short-time objective intelligibility.

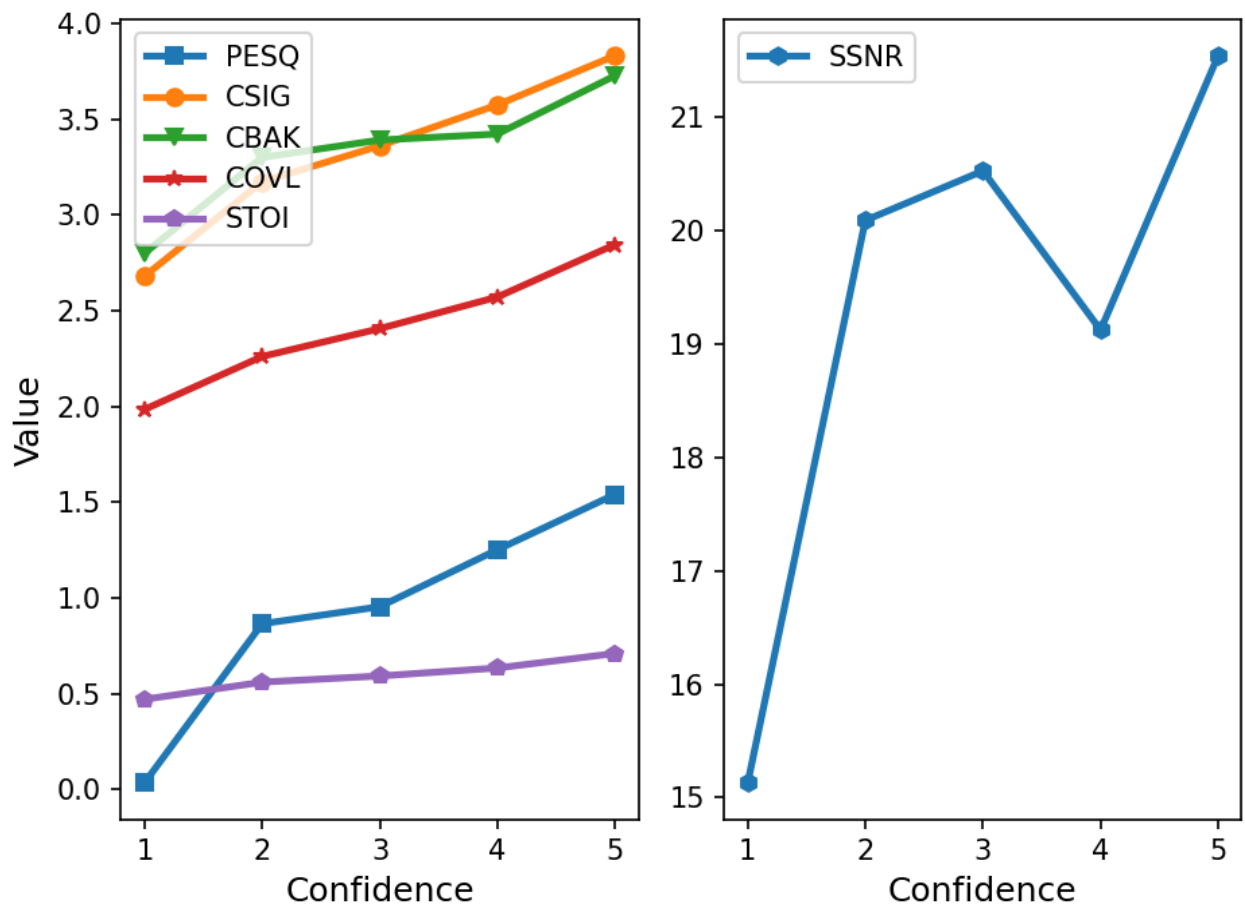

### Ablation Study

*Other Classification Model*

To assess the effectiveness of our proposed speech enhancement preprocessing technique with different classification models, we conducted an ablation study. The hyperparameters used in this study are detailed in Multimedia Appendix 2. We used the fine-tuned CMGAN as the speech enhancement module as it showed consistently outstanding performance in previous experiments, as shown in Table 2.

For the ICBHI dataset, the speech enhancement preprocessing technique increased the sensitivity by 11.71% and the ICBHI score by 1.4% when using the AST model [54]. Similarly, when using the AST model with the Patch-Mix strategy [16], the speech enhancement preprocessing technique increased the sensitivity by 17.08% and the ICBHI score by 1.6%, as shown in Tables 5 and 6.

Regarding the FABS dataset, the speech enhancement preprocessing technique increased the sensitivity by 18.48% and the ICBHI score by 5.46% when fine-tuning the AST model [54]. When fine-tuning the AST model using the Patch-Mix strategy [16], the speech enhancement preprocessing technique increased the sensitivity by 13.04% and the ICBHI score by 0.68%, as shown in Tables 7 and 8.

These results demonstrate that the speech enhancement preprocessing technique effectively improves the performance of various respiratory sound classification models, including fine-tuning the AST and AST using the Patch-Mix strategy, on both the ICBHI and FABS datasets.

**Table 5.** Comparison of the classification performance on the International Conference in Biomedical and Health Informatics (ICBHI) database by fine-tuning the Audio Spectrogram Transformer [54].

|  | Accuracy (%) | Sensitivity (%) | Specificity (%) | ICBHI score (%) |
| --- | --- | --- | --- | --- |
| Clean | 70.65 | 64.88 | 75.67 | 70.27 |
| Noisy | 24.13 | 30.41 | 18.67 | 24.54 |
| Noise injection | 53.78 | 35.28 | 69.87 | 52.58 |
| Audio enhancement | 54.46 | 46.99 | 60.96 | 53.98 |

**Table 6.** Comparison of the classification performance on the International Conference in Biomedical and Health Informatics (ICBHI) database using the Patch-Mix training strategy from the Audio Spectrogram Transformer pretrained weight [16].

|  | Accuracy (%) | Sensitivity (%) | Specificity (%) | ICBHI score (%) |
| --- | --- | --- | --- | --- |
| Clean | 70.73 | 61.79 | 78.5 | 70.14 |
| Noisy | 29.05 | 35.45 | 23.48 | 29.46 |
| Noise injection | 58.02 | 23.9 | 87.69 | 55.8 |
| Audio enhancement | 58.55 | 40.98 | 73.83 | 57.4 |

**Table 7.** Comparison of the classification performance on the Formosa Archive of Breath Sound database by fine-tuning the Audio Spectrogram Transformer [54].

|  | Accuracy (%) | Sensitivity (%) | Specificity (%) | ICBHI[a] score (%) |
| --- | --- | --- | --- | --- |
| Clean | 85.74 | 46.74 | 94.21 | 70.48 |
| Noisy | 83.03 | 36.96 | 93.03 | 65 |
| Noise injection | 83.8 | 31.52 | 95.16 | 63.34 |
| Audio enhancement | 80.89 | 50 | 87.6 | 68.8 |

[a]ICBHI: International Conference in Biomedical and Health Informatics.

**Table 8.** Comparison of the classification performance on the Formosa Archive of Breath Sound database using the Patch-Mix training strategy from the Audio Spectrogram Transformer pretrained weight [16].

|  | Accuracy (%) | Sensitivity (%) | Specificity (%) | ICBHI[a] score (%) |
| --- | --- | --- | --- | --- |
| Clean | 86.13 | 42.39 | 95.63 | 69.01 |
| Noisy | 82.15 | 29.35 | 93.62 | 61.49 |
| Noise injection | 82.44 | 44.57 | 90.67 | 67.62 |
| Audio enhancement | 75.17 | 57.61 | 78.98 | 68.3 |

[a]ICBHI: International Conference in Biomedical and Health Informatics.

### Metric Discriminator

Given that the metric discriminator optimizes PESQ, a metric primarily used in the speech domain for speech quality, a potential mismatch problem may arise when applied to respiratory sound tasks. To explore this issue, we conducted ablation studies on CMGAN's discriminator, examining the conformer generator-only model, the conformer generative adversarial network without PESQ estimation discriminator (with normal discriminator), and the complete setup (with metric discriminator). As shown in Table 9, the addition of a metric discriminator improved overall accuracy, sensitivity, and ICBHI score. This outcome indicates a positive contribution of the metric discriminator on PESQ to respiratory sound classification.

**Table 9.** Classification results of the convolution-augmented transformer–based metric generative adversarial network with different discriminator setups on the International Conference in Biomedical and Health Informatics (ICBHI) 2017 database.

| Setup | Accuracy (%) | Sensitivity (%) | Specificity (%) | ICBHI score (%) |
| --- | --- | --- | --- | --- |
| Generator only | 65.81 | 58.21 | 72.42 | 65.32 |
| With normal discriminator | 66.19 | 55.61 | 75.39 | 65.5 |
| With metric discriminator | 66.72 | 62.28 | 70.58 | 66.43 |

## Discussion

### Principal Findings

This paper proposes a deep learning audio enhancement preprocessing pipeline for respiratory sound classification tasks. We also introduced a collection of clinical noise and a real-world respiratory sound database from the emergency department of the Hsin-Chu Branch at the NTUH. Our noise-robust method enhances model performance in noisy environments and provides physicians with improved audio recordings for manual assessment even under heavy noise conditions.

The experimental results indicated that audio enhancement significantly improved performance across all 3 types of noise commonly encountered during auscultation. Specifically, our approach achieved a 2.15% improvement ($P<.001$) over the conventional noise injection method on the ICBHI dataset and outperformed it by 2.79% ($P<.001$) on the FABS dataset. Moreover, time-frequency–domain enhancement techniques demonstrated superior performance for this task. Analyzing the correlation between classification performance and speech quality metrics, we observed that CBAK and SSNR exhibited higher correlations with ICBHI scores. These metrics are strongly influenced by background noise but are unrelated to speech intelligibility, aligning with the experimental settings. In the physician validation study, enhanced recordings showed an 11.61% increase in sensitivity and a 14.22% improvement in classification confidence. A positive correlation was also observed between speech quality metrics and diagnostic confidence, highlighting the effectiveness of enhanced recordings in aiding physicians in detecting abnormal respiratory sounds. Our ablation study on various classification model structures revealed that audio enhancement preprocessing consistently improved performance. The findings showed enhanced sensitivity and higher ICBHI scores across both databases when tested with 2 state-of-the-art respiratory sound classification models. Furthermore, incorporating the metric discriminator PESQ was found to enhance downstream classification performance.

These findings validate the feasibility and effectiveness of integrating deep learning–based audio enhancement techniques into respiratory sound classification systems, addressing the critical challenge of noise robustness and paving the way for the development of reliable clinical decision support tools.

## Limitations and Future Work

Despite the encouraging findings in this study, there is a need to explore the co-optimization of front-end audio enhancement and classification models. As most audio enhancement tasks primarily focus on speech, the evaluation metrics are not highly correlated with respiratory sounds, potentially leading to inefficient optimization. Addressing this issue is crucial for achieving better performance in respiratory sound classification in future work. Furthermore, future studies should incorporate other types of noise and more complex noise mixture strategies to enable the development of a more noise-robust respiratory sound classification model for real-world clinical use. By considering a diverse range of noise scenarios, the model can be better prepared to handle the variability and challenges encountered in actual clinical settings. In addition, we have to speed up the model inference by simplifying the model to make it suitable for real-time applications. At the same time, we must ensure that enhancement quality is maintained and critical respiratory sound characteristics are preserved. In our long-term future work, we aim to deploy this model in real clinical environments by integrating it into electronic stethoscopes. To ensure the method's generalizability, we plan to collect cross-site respiratory sound recordings from 100 patients across various clinical environments. Of these recordings, data from 80 patients will be used for training, whereas data from the remaining 20 patients will be reserved for testing as part of a validation process aligned with Food and Drug Administration requirements. This approach will help validate the model's performance and facilitate its adoption for practical use in clinical settings.

## Conclusions

In this study, we investigated the impact of incorporating a deep learning–based audio enhancement module into automatic respiratory sound classification systems. Our results demonstrated that this approach significantly improved the system's robustness and clinical applicability, particularly in noisy environments. The enhanced audio not only improved classification performance on the ICBHI and FABS datasets but also increased diagnostic sensitivity and confidence among physicians. This study highlights the potential of audio enhancement as a critical component in developing reliable and trustworthy clinical decision support systems for respiratory sound analysis.

### Acknowledgments

This research is funded by the National Science and Technology Council of Taiwan under grant 112-2320-B-002-044-MY3.

### Conflicts of Interest

None declared.

### Multimedia Appendix 1

Details of the technical setup used in this study.
[DOCX File , 17 KB-Multimedia Appendix 1]

### Multimedia Appendix 2

Hyperparameters for training enhancement and classification models.
[DOCX File , 20 KB-Multimedia Appendix 2]

## Abbreviations

**AI:** artificial intelligence
**AST:** Audio Spectrogram Transformer
**CBAK:** mean opinion score of background noise intrusiveness
**CMGAN:** convolution-augmented transformer–based metric generative adversarial network
**CNN:** convolutional neural network

**CNN14:** 14-layer convolutional neural network
**Conformer:** convolution-augmented transformer
**FABS:** Formosa Archive of Breath Sound
**ICBHI:** International Conference in Biomedical and Health Informatics
**MOS:** mean opinion score
**NTUH:** National Taiwan University Hospital
**PESQ:** perceptual evaluation of speech quality
**PHASEN:** Phase-and-Harmonics–Aware Speech Enhancement Network
**SNR:** signal-to-noise ratio
**SSNR:** segmental signal-to-noise ratio
**STFT:** short-time Fourier transform
**STOI:** short-time objective intelligibility
**SVM:** support vector machine